\documentclass[ reprint, amsmath,amssymb, aps,]{revtex4-2}

\usepackage{graphicx}
\usepackage{dcolumn}
\usepackage{bm}
\usepackage{graphicx}

\usepackage{xcolor}

\usepackage{dsfont}

\begin{document}

\title{Spacetime Dynamics and Local Entropy Balance on Causal Horizons}

\author{Daegene Song}
\affiliation{Department of Management Information Systems, Chungbuk National University, Cheongju, 28644 Korea.}

\date{\today}

\begin{abstract}

We propose that spacetime dynamics can be organized by a Planck-scale bookkeeping rule, written using a modular-parameter normalization of size $2\pi$, that balances the geometric entropy increment $\delta A/4G$ against a reversible modular-energy flow $\delta\!\langle K\rangle$ and an irreversible Landauer--Bennett cost $\ln 2\,\delta N_c$, where $K\equiv K_\sigma$ is the (dimensionless) modular Hamiltonian of the chosen region defined relative to a fixed reference state $\sigma$, and $N_c$ counts logically irreversible one-bit record updates (e.g. coarse-grained overwrites/registrations) on that screen. This ``information--geometry ledger'' is consistent with the Bekenstein--Hawking area law, and—when enforced on small causal screens under the standard entanglement-equilibrium assumptions—recovers the full nonlinear Einstein equation. In FLRW cosmology, the same bookkeeping motivates a two-component vacuum sector $\rho_{\rm vac}=\rho_\Lambda+3\varepsilon H^{2}/8\pi G$ when a constant inefficiency parameter $\varepsilon$ is assumed.

\end{abstract}

\maketitle

\section{Introduction}

Over the past five decades, a converging set of ideas has linked gravitation to information. Black-hole thermodynamics revealed an area–entropy relation \cite{bekenstein1973} and a thermal character for horizons \cite{hawking1975,gibbons1977}, while quantum-information tools—most notably the first law of entanglement \cite{blanco2013,lashkari2014} and modular Hamiltonians \cite{faulkner2016}—have clarified how small deformations of states and regions encode energetic and entropic responses \cite{wald1994}. In parallel, ``entanglement-equilibrium'' arguments have shown that, in suitable local setups, gravitational field equations follow from enforcing informational balance on small causal diamonds \cite{jacobson1995, jacobson2016}, without committing to detailed microphysics \cite{padmanabhan2010}. Despite this progress, the literature oscillates between global and local viewpoints \cite{bousso2002}, reversible and irreversible processes \cite{unruh1982}, and model-dependent assumptions about matter and horizon structure. What is missing is a compact, operational statement—expressed directly in entropy units—that an individual observer can apply on arbitrarily small screens, independent of microscopic details, and that cleanly interfaces with known thermodynamic \cite{callen1985} and information-theoretic constraints \cite{shannon1948}.

A second motivation comes from dynamical horizons and cosmology. Apparent horizons in FLRW spacetimes carry temperature and entropy in close analogy with their black-hole counterparts \cite{hayward1998, bak2000, cai2005}, yet they evolve in time and thus invite a careful accounting of both reversible entanglement flow \cite{casini2008} and genuinely irreversible record formation \cite{landauer1961}. The latter has a well-defined informational cost through Landauer’s principle, suggesting that any local gravitational bookkeeping should distinguish, and consistently balance, these two channels \cite{bennett2003}. Such a perspective is also natural in phenomenological contexts where vacuum contributions may track curvature scales \cite{weinberg1989}. These considerations motivate the framework developed below, which formulates a minimal, observer-based information accounting and then examines its consequences for gravitational dynamics and cosmology \cite{frieman2008}.

Throughout the paper we therefore treat the local entropy-balance ``ledger'' relation, introduced in Sec.~\ref{local}, not as an independent dynamical postulate but as a compact reformulation of well-established ingredients: the Bekenstein–Hawking area–entropy relation, the entanglement first law for small deformations of states and regions, and Landauer’s principle for classical records. The recovery of Einstein’s equation and the emergence of a two-component vacuum sector in FLRW cosmology are then derived consequences of applying this balance in different local settings rather than additional assumptions.

\section{Local Postulate}
\label{local}

At each proper-time instant along an observer’s worldline, select a small, compact spacelike two-surface (a ``cut'') $\mathcal H$ that partitions a Cauchy slice into a chosen region accessible to the observer. Let $\sigma$ be a fixed reference state on the degrees of freedom of that chosen region, and let $K_\sigma$ denote the corresponding (dimensionless) modular Hamiltonian, so that the entanglement first law gives $\delta S_{\rm ent}=\delta\!\langle K_\sigma\rangle$ for small state/shape deformations. For concreteness, one may picture $\mathcal H$ as the spherical intersection of the future light cone of an event $p$ with a small Cauchy slice, or equivalently as a small patch of a local Rindler horizon whose exterior region is accessible to the observer.

We frame variations using a modular-parameter interval of size $2\pi$, i.e. the standard KMS normalization used in local Rindler/CHM equilibrium setups. We do not assume that modular flow is periodic in general; ``$2\pi$'' here is a normalization convention for the modular parameter in the local equilibrium regimes considered below. For any infinitesimal physical deformation of the cut during such a lap, the following dimensionless balance holds:
\begin{equation}
\frac{1}{4G}\,\delta A_{\mathcal H}
\;=\;
-\,\delta\!\bigl\langle K_\sigma\bigr\rangle
\;+\;
\ln 2\,\delta N_c,
\label{eq:ledger}
\end{equation}
where $A_{\mathcal H}$ is the area of the cut in the physical metric $g_{\mu\nu}$, $\langle K_\sigma\rangle$ is the modular-energy expectation for the actual state of the chosen region, and $N_c$ counts logically irreversible classical one-bit record updates on $\mathcal H$ during the interval (e.g. coarse-grained overwrites/registrations) \cite{landauer1961,bennett1982}. By definition $\delta N_c\!\ge\!0$, and in reversible deformations $\delta N_c\!=\!0$.  We use $c=\hbar=k_B=1$.

Restoring \(c\) and \(\hbar\) (while still taking \(k_B=1\)), the same balance reads
\begin{equation}
\frac{c^{3}}{4G\hbar}\,\delta A_{\mathcal H}
\;=\;
-\,\delta\!\bigl\langle K_\sigma\bigr\rangle
\;+\;
\ln 2\,\delta N_c,
\label{eq:ledger-dimful}
\end{equation}
since $K_\sigma$ and $N_c$ are dimensionless, so that $c^{3}A_{\mathcal H}/G\hbar$ is also dimensionless. Thus \eqref{eq:ledger} is a manifestly dimensionless statement relating three contributions to the entropy budget.

Equation \eqref{eq:ledger} rests on three standard ingredients: (i) the entanglement first law $\delta S_{\rm ent}=\delta\!\langle K_\sigma\rangle$ for small deformations of the state and the cut, (ii) the identification of $A_{\mathcal H}/4G$ with the Bekenstein–Hawking entropy associated with the screen, and (iii) Landauer’s bound $Q\ge k_B T\ln 2$ per irreversibly erased classical bit, which we encode via the integer $N_c$ of logically irreversible updates on $\mathcal H$. The ledger simply asserts that, per modular $2\pi$ interval, the change in geometric entropy $A_{\mathcal H}/4G$ is compensated by a reversible modular-energy flow and an irreversible record-keeping cost.

\section{Bekenstein–Hawking Entropy}
\label{BHcheck}
We now check that the ledger \eqref{eq:ledger}, with its geometric normalization $1/4G$, is consistent with the Bekenstein–Hawking area law for stationary black holes. Consider a quasi-static family of nearby stationary black holes (Schwarzschild or Kerr) with no work terms; the process is reversible and we set $\delta N_{\rm c}=0$. Here $M$ denotes the ADM mass and $\kappa$ the surface gravity of the Killing horizon. In this static setting the reference state $\sigma$ is taken to be the Hartle–Hawking (or appropriate stationary) state. For the choice where the exterior is the chosen region, the modular first-law relation gives $\delta\!\langle K_\sigma\rangle=\beta_H\,\delta E_{\rm ext}$ with $\beta_H=1/T_H$. For a quasi-static infall that increases the black-hole mass by $\delta M$, the exterior energy decreases by $\delta E_{\rm ext}=-\delta M$, hence $\delta\!\langle K_\sigma\rangle=-\beta_H\,\delta M$. The mechanical first law gives
\[
\delta M \;=\; \frac{\kappa}{8\pi G}\,\delta A \,,
\]
while reversibility implies $\delta M=T_H\delta S_{\rm grav} = (\kappa/2\pi) \delta S_{\rm grav}$. These two relations are compatible if and only if $\delta S_{\rm grav} \;=\; \frac{1}{4G}\,\delta A $,
so integrating and fixing $S_{\rm BH}(A\!=\!0)=0$ yields the familiar area law
\[
S_{\rm BH}(A)\;=\;\frac{A}{4G}\,.
\]
Thus, the geometric term in \eqref{eq:ledger} is normalized so that, in the reversible sector with $\delta N_{\rm c}=0$, it exactly reproduces the black-hole entropy and the reversible part of the ledger reduces to the standard relation between modular energy and gravitational entropy for stationary horizons.

\section{Field Equation}
The information-geometry ledger can be pushed one step further: by enforcing it on an infinitesimal causal screen that encloses a tiny laboratory, we recover the full nonlinear Einstein field equations \cite{jacobson2016}. The key idea is to regard the screen as the boundary of a geodesic ball of radius $R$ centered at an event $p$. When $R$ is much smaller than any curvature scale, spacetime inside the ball is almost Minkowskian, so quantum fields behave as if they lived in flat space while geometry records only small, quadratic curvature corrections. In this section we take the chosen accessible region to be the geodesic ball $B_R$ itself, so the modular Hamiltonian used in the ledger is $K_\sigma=K_{B_R}$ for $\sigma$ the vacuum (or appropriate local KMS reference) restricted to the algebra of $B_R$.

Working in Riemann normal coordinates at $p$, the metric reads $g_{\mu\nu}=\eta_{\mu\nu}+{\cal O}(R^{2})$. To leading order, the only geometric data that influence the ball are the components of the Riemann tensor.  On the quantum side, we adopt the standard entanglement-equilibrium/small-ball assumption: in the $R\to 0$ regime (and in particular for CFT vacuum, or more generally for theories with CFT-like UV behavior), the vacuum modular Hamiltonian for a ball is well-approximated by the Casini–Huerta–Myers (CHM) form \cite{casini2011}
\begin{equation}
K_{B_R}= 2\pi \int_{B_R}\!\mathrm d^{3}x\,\frac{R^{2}-r^{2}}{2R}\,T_{00}(x).
\end{equation}
The first law of entanglement \cite{blanco2013,lashkari2014,faulkner2016} states that, for small perturbations of the state and region, the variation of entanglement entropy equals the variation of the expectation value of the modular Hamiltonian,
\begin{equation}
\delta S_{\text{ent}}=\delta\langle K_{B_R}\rangle.
\end{equation}
Evaluating the right-hand side with the CHM kernel for a perturbation localized near $p$ yields
\begin{equation}
\delta S_{\text{ent}}=\frac{8\pi^{2}}{15}\,R^{4}\,\Delta T_{00}(p),
\end{equation}
where $\Delta T_{00}(p)$ is the change in local energy density relative to the same reference state $\sigma$ used to define $K_{B_R}$.

The geometric side of the ledger requires the area of the ball’s boundary at fixed volume. Jacobson showed in \cite{jacobson2016} that curvature reduces this area by
\begin{equation}
\delta A_{|V}=-\frac{4\pi R^{4}}{15}\,G_{00}(p),
\end{equation}
with $G_{00}=R_{00}-\tfrac12Rg_{00}$ the $00$ component of the Einstein tensor. The ledger equates the entanglement entropy flux to the gravitational entropy change, and because the cut is taken reversible we set $\delta N_{\rm c}=0$. Using $\delta S_{\text{grav}}=\delta A/(4G)$ we find
\begin{equation}
-\frac{\pi R^{4}}{15G}\,G_{00}(p)+\frac{8\pi^{2}}{15}\,R^{4}\,\Delta T_{00}(p) = 0
\label{G00}
\end{equation}

To promote this single component to a full tensor equation, define $H_{\mu\nu}\equiv G_{\mu\nu}-8\pi G\,T_{\mu\nu}$. Here $T_{\mu\nu}$ may be taken as the stress-tensor expectation value relative to the same reference state $\sigma$ used above (so $T_{00}=\Delta T_{00}$ in Eq.~\eqref{G00}); equivalently, one may use the renormalized expectation value and absorb any $\sigma$-dependent subtraction into the integration constant $\Lambda$. The preceding argument applied in a local inertial frame with four-velocity $u^\mu$ at $p$ gives $H_{\mu\nu}u^{\mu}u^{\nu}=0$ for that choice of $u^\mu$. Because both $p$ and the orientation of the local inertial frame are arbitrary, repeating the construction shows that $H_{\mu\nu}u^{\mu}u^{\nu}=0$ for every unit timelike vector $u^{\mu}$. A standard linear-algebra lemma then implies that a symmetric tensor with vanishing contraction along all unit timelike vectors must be pure trace, i.e. $H_{\mu\nu}=\Phi g_{\mu\nu}$ for some scalar function $\Phi(p)$; no other symmetric tensor satisfies this condition at a point \cite{jacobson1995}. Taking a covariant divergence and using both $\nabla^\mu G_{\mu\nu}=0$ (Bianchi identity) and $\nabla^\mu T_{\mu\nu}=0$ (local energy–momentum conservation) forces $\nabla_\nu\Phi=0$, so $\Phi=-\Lambda$ is a spacetime constant. The result is the complete, nonlinear field equation \cite{jacobson1995,blanco2013} $G_{\mu\nu}+\Lambda\,g_{\mu\nu}=8\pi G\,T_{\mu\nu}$.

Several observations underscore the informational nature of this derivation. First, the coefficient $8\pi G$ is fixed by matching entropies; there is no adjustable coupling. Second, curvature enters only through the entropic cost of distorting a causal screen, while matter enters solely through the modular energy that measures how entanglement changes when the state is perturbed. Finally, the cosmological constant appears as an integration constant that preserves the entropy balance—its value is not supplied by the ledger but must be set by boundary conditions or experiment.

In this light, Einstein’s equation is an equilibrium condition stating that to first order in $R$ every geodesic ball is poised at entanglement equilibrium: any attempt to raise or lower modular energy must be accompanied by an equal and opposite geometric entropy shift. Gravity therefore emerges not as a fundamental interaction but as a bookkeeping rule that guarantees local consistency between the quantum information stored in fields and the information encoded in spacetime geometry.

The derivation crucially exploits the universality of short-distance entanglement. Because the CHM kernel holds exactly in CFT vacuum and is expected to approximate the modular Hamiltonian in the small-ball limit for theories with sufficiently CFT-like UV behavior (up to corrections suppressed by $R/\ell_{\rm UV}$ and by relevant deformations), the argument remains agnostic about the particle spectrum, couplings, or even whether the underlying theory is conformal at macroscopic scales. Conversely, any modification of quantum theory that alters the first law of entanglement would necessarily modify the left-hand side of the ledger and hence the form of the gravitational field equations, providing a sharp diagnostic for beyond-quantum proposals.

\section{Running Vacuum}
The cosmic acceleration problem provides an ideal playground for the information–geometry ledger \eqref{eq:ledger}. In contrast to the stationary black-hole case, a FLRW universe features a dynamical apparent horizon whose area changes with the Hubble rate \cite{hayward1998}. Each time slice therefore induces both reversible entanglement flow, encoded in the modular-energy term $\delta\!\langle K_\sigma\rangle$, and (potentially) irreversible classical information loss on the horizon, encoded in the $\delta N_c$ term. Because the ledger balances these two channels separately, it naturally predicts a vacuum energy sector with two additive pieces: an integration constant that survived the derivation of Einstein’s equation and a running contribution controlled by an inefficiency parameter $\varepsilon$ \cite{basilakos2009}.

For a spatially flat FLRW background with scale factor $a(t)$ the apparent-horizon radius is $R_A=1/H(t)$, where $H=\dot a/a$. The associated area and Gibbons–Hawking entropy are
\[
A=\frac{4\pi}{H^{2}}\; ,\; S=\frac{A}{4G}=\frac{\pi}{G\,H^{2}},
\]
while the horizon temperature is $T_H=H/2\pi$ \cite{gibbons1977}. These relations mirror their black hole counterparts but with $H(t)$ playing the role of an effective surface gravity \cite{gong2007}.

Differentiating $S$ gives an entropy production rate
\[
\dot S=-\frac{2\pi}{G}\,\frac{\dot H}{H^{3}}\,.
\]
(For the standard expanding decelerating $\Lambda$CDM-like regime one has $\dot H<0$, hence $\dot S>0$). Guided by the ledger \eqref{eq:ledger}, we split this total geometric entropy change into a reversible part associated with modular-energy flow and an irreversible contribution corresponding to the $\delta N_c$ logically irreversible classical bits recorded on the horizon. Writing $\dot S_{\mathrm{irr}}=(\ln 2)\,\dot N_c$ and parametrizing the ratio of irreversible to total entropy production by
\[
\varepsilon \;\equiv\; \frac{\dot S_{\mathrm{irr}}}{\dot S}
\]
we have $\dot S_{\mathrm{irr}}=\varepsilon \dot S$ and $\dot S_{\mathrm{rev}}=(1-\varepsilon)\dot S$, with $0\le\varepsilon\le 1$ measuring the fraction of horizon degrees of freedom that are genuinely reset per Hubble time. Landauer’s principle converts the irreversible part into a heat flux
\[
\Phi_Q=-T_H\,\dot S_{\mathrm{irr}}=\frac{\varepsilon\,\dot H}{G\,H^{2}}\,.
\]
This is the only phenomenological input in the cosmological application of the ledger: we assume that $\varepsilon$ is slowly varying and treat it as a constant on cosmological time-scales.

Energy conservation inside the spherical region of proper volume $V=4\pi/3H^{3}$ is governed by Hayward’s unified first law,
\[
\dot\rho+3H(\rho+p) = \frac{\Phi_Q}{V} = \frac{3\varepsilon H\dot H}{4\pi G}
\]
where $\rho$ and $p$ refer to the fluid enclosed by the horizon \cite{hayward1998}. For simplicity we assume the ordinary matter sector obeys its standard conservation law $\dot\rho_m+3H(\rho_m+p_m)=0$, and attribute the source term on the right-hand side to the vacuum-like component. If we separate a vacuum-like component $(\rho_\varepsilon,p_\varepsilon)$ from ordinary matter and assume it obeys $p_\varepsilon=w\rho_\varepsilon$, the above equation is solved by
\[
\rho_\varepsilon(t)=\frac{3\varepsilon H^{2}(t)}{8\pi G},\; w=-1.
\]
Thus, a constant inefficiency factor automatically generates a running vacuum density $\propto H^{2}$ while preserving the equation of state required for accelerated expansion.

The first-law derivation of Einstein’s equation already supplied an integration constant $\Lambda$, interpreted here as a density $\rho_\Lambda=\Lambda/8\pi G$. Combining the two pieces,
\[
\rho_{\mathrm{vac}}(t)=\rho_\Lambda+\frac{3\varepsilon H^{2}(t)}{8\pi G},
\]
one obtains a two-component vacuum sector in which $\rho_\Lambda$ is genuinely time-independent whereas the $\rho_\varepsilon$ term decays with the cosmic expansion \cite{basilakos2009}. Phenomenologically, running-vacuum fits typically constrain the corresponding parameter to be small (often quoted at the $\sim 10^{-3}$–$10^{-4}$ level in model- and dataset-dependent analyses), and small nonzero values can remain compatible with supernovae, baryon acoustic oscillations, and CMB data \cite{peracaula2023}.

From an informational perspective, $\rho_\Lambda$ stems from the reversible part of the ledger: it is the constant of integration required by the Bianchi identity once entanglement equilibrium is enforced on every causal diamond. By contrast, $\rho_\varepsilon$ measures the inefficiency of the cosmic horizon as an information engine: it is proportional to the rate at which irreversible classical bits, encoded in $\delta N_c$, are recorded on the horizon. Whenever new classical records are irreversibly written on the horizon, Landauer heat must be dumped, and the only available sink is the vacuum energy stored inside the horizon.

Keeping $\varepsilon$ constant maintains $w=-1$ and preserves the simplicity of the Friedmann equations. Allowing $\varepsilon(t)$ to vary would introduce additional source terms $\propto\dot\varepsilon$ in the continuity equation, spoiling the pure-vacuum equation of state and severely complicating phenomenology. Until observations demand otherwise, the constant-$\varepsilon$ ansatz therefore represents the minimal, information-theoretic extension of the cosmological constant paradigm.

\section{Conclusions and Outlook}
The information-geometry ledger proposed here recasts gravity as the local closure of an entropy balance sheet: the geometric increment $\delta A/4G$ on every Planck-pixel horizon element must be offset by reversible modular energy flow and the irreversible Landauer cost of classical record-keeping. Enforcing this bookkeeping aligns with the Bekenstein–Hawking area law, elevates the ledger to the full nonlinear Einstein equation, and predicts a two-component vacuum sector in FLRW cosmology. In this perspective, spacetime is not a fundamental arena but an adaptive ledger that enforces informational equilibrium between quantum fields and geometry.

The informational origin of gravity suggested here invites concrete empirical probes. In the laboratory, indirect tests of the entanglement first law—via relative-entropy/entanglement-temperature measurements in cold-atom or superconducting-circuit platforms—can bound deviations from $\delta S=\delta\!\langle K\rangle$. Separately, precision Landauer calorimetry constrains the irreversible term that motivates our $\varepsilon$ partition in FLRW. Cosmologically, next-generation surveys of type Ia supernovae, baryon acoustic oscillations, and the CMB could reach the few $10^{-4}$ level, subject to systematics and degeneracies. On the theoretical side, extending the ledger to far-from-equilibrium settings may illuminate black hole evaporation, singularity resolution, and the quantum-to-classical transition in the early universe. In all these contexts, the guiding principle remains the same:  geometry must balance information. Whether one approaches gravity from quantum information, thermodynamics, or cosmology, the ledger furnishes a unifying, parameter-free scaffold on which to build—and to test—an emergent description of spacetime.

\end{document}